# Refactoring Technical Support to Reduce Interrupts of Developers

*Zádor Dániel Kelemen, Balázs Tódor, Sándor Hodosi, Ákos Somfai*
NNG LLC., 1037 Budapest Szépvölgyi út 35, Hungary
{Daniel.Kelemen, Balazs.Todor, Sandor.Hodosi, Akos.Somfai}@NNG.com

**Abstract**

In this paper an analysis of a technical support data with the goal of identifying process improvement actions for reducing interrupts is presented.

A technical support chat is established and used to provide internal developer support to other development teams which use the software code developed by a core team. The paper shows how data analysis of a 6 months support time helped to identify gaps and action items for improving the technical support process to minimize interrupts from other developer teams.

The paper also shows effects (advantages and drawbacks) of refactor actions taken based on this analysis.

# 1  Introduction

Classical management and process improvement frameworks use measurement as a tool for continuous improvement. Upper levels of widespread SPI approaches such as CMMI or (Automotive) SPICE recommend quantitative management of work products and processes [1], [2]. These approaches focus on "what" to be done, and they intentionally do not provide further guidance on "how" to analyse data or "from where to collect useful data".

This paper shows how action items of a process improvement were identified at a core automotive software development team supporting a high number of projects (and high number of other developers) based on simple analyses of data extracted from a modern messaging application.

As a trigger of this analysis, management of a business unit at NNG LLC. requested a root cause analysis to investigate why technical support time was high at a core team. The initial root cause analysis was broken down to several sub-analyses, including the analysis of how interrupts of developers could be reduced. The scope of this paper is to focus only on reducing interruptions and on identifying process improvement opportunities related to the reduction of interruptions and to review effects of improvement actions taken. It is not in the scope to describe the whole root cause analysis performed at the core team.

NNG is a global leader of automotive navigation software, has 700+ employees in different locations, most of them in 5 different buildings in Budapest, Hungary. After several discussions with the management representatives and interviews with the technical staff it came out that one main channel of technical support requests is a Skype chat, having members from all over the organisation including the members of the investigated core team.

From development point of view support chat requests (messages) are considered interrupts. Drawback of developer time interrupts has been investigated by multiple researchers and it was shown that interrupts can have negative effects on software development performance and the cost and effects (e.g. recovery time after an interrupt) can be quantified [3], [4]. Due to the negative effects of interrupts it was considered relevant to perform the analysis and identify action items accordingly.

Section 2 describes the approach used in this paper, section 3 presents the data collection and data preparation, section 4 shows the data analysis on collected data, and section 5 provides a brief summary of improvements based on analysis results. Section 6 describes the effects of the improvement actions taken. The paper ends with limitations in section 7 and conclusion in section 8 respectively.

# 2  Approach

The question to be answered in this paper is: "How interruptions from technical support chat could be reduced based on chat log analysis?"

In order to answer the question the following steps were identified:
   I. Interrupt data collection and data preparation (discussed in 3),
   II. Interrupt data analysis (discussed in 4),
   III. Identification of process improvement actions based on data analysis results (discussed in 5),
   IV. Discussion of the effects of process improvement actions taken (included in 6).

# 3  Data collection and data preparation

In order to perform the organisation-related data analysis of a support chat, two data types were identified: (1) the logs of the chat and (2) organisation related data such as roles and composition of teams.

The following steps were performed during the data collection and data preparation phase:

1. Data collection from support chat

   A 6 months chat log was provided by the team leader involving 140+ active days which was considered sufficient for the analysis. Data from Skype has been collected by making use of SkypeLogView tool [5].

2. User data collection from internal database

   At NNG, list of employees, teams and various contact information including Skype are stored in a database. Skype IDs of active users of the support chat were used as a search key in the internal database.

3. Data preparation

   Data preparation consisted of merging user information collected from the internal database with the data collected from support chat log. All activities were considered as message sending (broadcasting) and end-line characters were removed from multi-line messages (considered as one message). Messages sent with the same timestamp by the same user were considered as single, multi-line messages.

## 4 Data analysis

According to the SPI Manifesto [6], involvement of the people started at the very beginning, metric candidates were identified in a brainstorming: total number of users, number of active users in a period, number of inactive users in a period, total messages, number of messages of the investigated team, external messages, total messages per user (most and less active users), total messages per role (most active role), average number of messages / day, average number of messages / hour and conversation length.

A period of **~6 months** has been analysed (2014.7.7-2015.2.11), including **218** days of which **144** were active days having 3529 messages in total. A day is considered active when at least one message is sent. Only partial data were available on the first and last days, therefore in some of the analyses these two days were excluded, taking into account only 142 active days (e.g. when calculating daily averages) with 3498 messages in total.

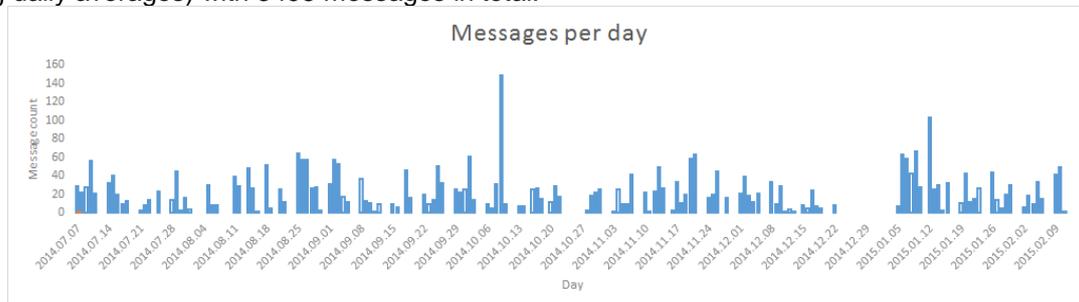

**Figure 1 – Distribution of messages per day**

The cost/benefit ratio of a measurement is always a central question. Therefore it was decided that only quick and simple measurements will be performed (and not all the possible analyses). For example "conversation length" was excluded because it could be difficult to measure real length of a conversation when multiple users interact in the chat room. Skype chats are working in a broadcasting mode: all members get all messages. Thus, it is difficult to identify attributes of conversations such as start time, end time, interrupts by next conversation start, all those which are needed to identify conversation lengths.

Taking into the account the metrics cost/benefit ratios the following set of the metrics were identified to be measured: messages per day, hourly distribution of messages, distribution of messages per weekdays, active versus inactive users, activeness of teams, activeness of roles and behaviours of top active users. In this section these metrics are discussed.

### 1. Messages per day

In order to identify peak days, message distribution per active days was checked first.

Figure 1 shows the message distribution for the entire analysed period. The days with most messages had line counts of 149, 103, 67, 64, 63, 63, 61, 59, 59 and 58. For further information on the average and median of messages sent per day in this period see Table 4.

### 2. Hourly distribution of messages

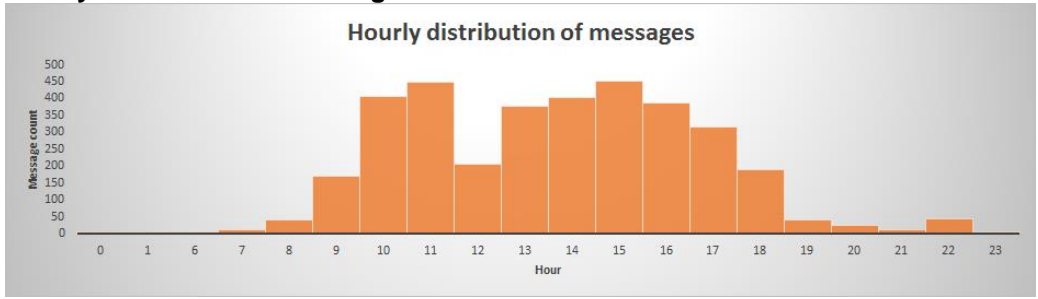

**Figure 2 – Hourly distribution of messages**

Measuring the peak hours of support time was in the scope of investigation. Figure 2 and Table 1 show the hourly distribution of messages. Peak hours are **15-16** and **11-12**.

**Table 1 – Message distribution per hour**

| Hour | Messages | Average |
|------|----------|---------|
| 15   | 452      | 3,18    |
| 11   | 449      | 3,16    |
| 10   | 405      | 2,85    |
| 14   | 402      | 2,83    |
| 16   | 388      | 2,73    |

### 3. Peak days

Similarly to peak hours it was investigated if there were peak days. Messages were distributed among weekdays as follows: Monday (709), Tuesday (713), Wednesday (586), Thursday (817), Friday (621), Saturday (49) and Sunday (3). It can be seen that (1) some messages were sent on weekends and (2) *there are no considerable differences among the number of messages sent on working days.*

### 4. Active and inactive users

Since messages are broadcasted, it was interesting to see what percentage of users were inactive in the period analysed. 23% of users (35 out of 154) were inactive in the analysed period. NNG has 700+ employees meaning that 22% of employees are member of this support chat of which 17% of all employees were active in the investigated ~6 months.

### 5. Activeness of teams

The analysis of activeness of teams showed, that out of the 3529 messages, 1154 (33%) were sent by the investigated team (9 people out of 18 team members) and the rest were sent by other business units and teams. The analysis also showed that even business units with up to 30 active members sent less messages.

### 6. Activeness of roles

It was also in the scope of investigation to see if employees with the right roles (developers) are sending messages on the chat. It came out that the majority of messages (68%) are sent by developers.

Distribution of messages per role was as follows: Developer (2386 messages sent, 68,21% of total messages), Project Manager (187, 5,35%), Team Leader (176, 5,03%), UI Developer (165, 4,72%), Technical Lead (131, 3,74%), Software Tester (87, 2,49%), Test Automation Engineer (65, 1,86%), Product Engineer (61, 1,74%), Architect (45, 1,29%), Other roles (195, 5,57%).

### 7. Top active users

Table 2 shows activeness of top 10 most active users based on message count. It can be seen that the most active user sent 315 messages in total which resulted in only 2,22 average messages per day. It also can be seen that top 4 active users are members of the investigated team.

**Table 2 – List of top 10 active users**

| User Alias | Total messages | Avg. message / day | Team | Role |
|---|---|---|---|---|
| User 1 | 315 | 2,22 | Investigated team | Developer |
| User 2 | 293 | 2,06 | Investigated team | Developer |
| User 3 | 183 | 1,29 | Investigated team | Developer |
| User 4 | 141 | 0,99 | Investigated team | Developer |
| User 5 | 140 | 0,99 | External team | Developer |
| User 6 | 111 | 0,78 | External team | Team leader |
| User 7 | 107 | 0,75 | External team | Developer |
| User 8 | 98 | 0,69 | External team | Developer |
| User 9 | 93 | 0,65 | External team | Developer |
| User 10 | 89 | 0,63 | External team | Developer |

## 5 Improvements identified based on data analysis

Table 3 shows the ID and metric number (column 1), results deducted from Skype chat log analysis (column 2) and gaps identified (column 3).

**Table 3 – Analysis results and gaps**

| ID (metric) | Analysis result | Gap |
|---|---|---|
| S-F1 (1) | There are days on which support chat interaction is high. In these days it is common that 20+ messages are sent in peak hours, often resulting in an interrupt in every ~3 minutes (cases were checked). In case if users are listening, their day is practically lost on peak support days. | Developers are not protected from interrupts |
| S-F2 (1, 3) | It is unexpected when a peak support day occurs (no trend can be derived). | Developers are not protected from interrupts |
| S-F3 (2) | Peak support hours overlap peak developer hours (core office hours are between 10-16). | Developers are not protected from interrupts |
| S-F4 (6) | 68% of interaction is by developers. | No support role exist for support tasks |
| S-F5 (observed, not in 4) | It is difficult to search in (Skype) support chat log and new users have no access to the skype log, same questions may happen in future. | There is no knowledge base |
| S-F6 (5, 6,7) | Top commenters send 0,6-2,2 messages per day and the investigated team members send 0,7 messages per day in average, there is no continuous need for all developers to listen the support chat. | Developers are not protected from interrupts (while they could be protected!) |
| S-F7 (4) | There are 154 users of the support chat. 23% of users were inactive in the last 6 months. Many of them may be interrupted, especially during peak support days (they may delete or mute support chat to avoid interrupts). | No support chat mute guide |

Analysis results and gaps served as an input to the investigated team and to the quality management to identify process improvement opportunities (action items with responsibles and deadlines). Not all of them can be listed within the frame of this paper due to confidentiality reasons. However, most important ones (which can also be shared publicly) were: protect developers from interrupts by (1) defining a dispatcher service policy (2) with a weekly rotating dispatcher role, (3) developing and maintain-ing a knowledge base (FAQ page) to reduce the number of interrupts of the dispatcher and developers and (4) defining and institutionalizing a support chat mute guide for inactive users (with chat message keywords for activation). Another option for dispatching could be a non-rotating, full time dispatcher, however considering the average number of messages per day, dispatcher rotation was chosen.

## 6 Effects of actions taken

Effects of actions taken were assessed after 9 weeks of operation in the new settings (1) on a one-hour refactor retrospective meeting and (2) by measuring changes between investigated periods and measuring dispatcher activity.

*Refactor retrospective* Team members, the team leader and the quality manager attended the refactor retrospective meeting (in total 8 participants). The meeting focused on positive and negative aspects of refactor.

**Positive aspects:** According to the team feedback (1) the improvement project reached its goal: majority of the team members can work without interrupts, (2) a feeling of success - team members respond and others thank their service, (3) there is always one dispatcher, others do not need to continuously watch the support chat, (4) positive feedback from other teams, (5) since there is always somebody dispatching, response time decreased.

**To be improved:** (1) the team created a FAQ page, but its development is slow, (2) it is difficult for the dispatcher to find who the expert is in an area, (3) requests are not rated (e.g. based on urgency), (4) the dispatchers wish to solve all issues instantly, even if it is not needed by the policy, (5) some of the externals want bug fixes within the frame of support, (6) not everyone is suitable for the dispatcher role, (6) there are lots of meetings which the dispatchers has to attend and thus they has to be substituted on the support channel, (7) not all the team members use the mute guide.

Team members were also asked to estimate how much time the dispatching role requires in a week. The answer of 7 team members serving in dispatching role varied between 0,5 hours to 8 hours per week (0,5h; 1h; 1-2h; 3-4h; 3-5h; ~5h; 5-8h).

*Measuring changes between investigated periods and measuring dispatcher activity*

There was an increase in the average of support chat messages in the second investigated period. Table 4 shows a summary of major changes. These changes may be caused by external factors (e.g. organisational growth). Despite the increase in the number of messages and due to the refactor actions taken, team members experienced a decrease in time spent on support.

**Table 4 – Comparison of major metrics in the investigated periods**

|  | **Both periods** | **14.7.8-15.2.10** | **15.3.2-15.5.3** |
|---|---|---|---|
| Total messages | 4973 | 3498 | 1475 |
| Maximum nr of messages per day | 149 | 149 | 99 |
| Average nr of messages (all days) | 17.70 | 16.05 | 23.41 |
| Average nr of messages (active days) | 27.02 | 24.63 | 34.30 |
| Median of messages (all days) | 10 | 9 | 15 |
| Median of messages (active days) | 21 | 20 | 28 |
| All days investigated | 281 | 218 | 63 |
| Active days | 185 | 142 | 43 |
| Zero message days | 96 | 76 | 20 |
| Non-workdays | 90 | 70 | 20 |

Table 5 shows message statistics of a 9 week "dispatcher-enabled" period in a weekly breakdown, columns are: week number, date interval, all number of messages sent, messages sent by the investigated team, number of messages sent by others (externals to the team), percentage of messages sent by the team, number of messages sent by the dispatcher, number of messages sent by non-dispatcher members of the investigated team, percentage of the messages sent by dispatcher (vs other team members) and dispatcher id. It can be seen that instead of 33% (first period), 57% of the messages were sent by the investigated team. Furthermore, ~49% of team messages were sent by dispatchers. Substitutions (when dispatchers had to attend meetings) are not counted. Some team members were serving as a dispatcher multiple times (see the last column), area expert dispatchers were providing more direct answers (e.g. dispatcher 2), while dispatchers with lower domain experience were asking the help of the team members (e.g. dispatcher 3).

Rest of the initially identified metrics were not re-measured, since no major conclusion would be drawn on the effects of the refactor.

**Table 5 – Dispatcher activity in a weekly breakdown**

| Week | Date interval | All | Team | Others | Team (%) | Disp | Not disp | Disp (%) | Disp |
|---|---|---|---|---|---|---|---|---|---|
| W1 | 03.02 - 03.08. | 52 | 23 | 29 | 55.77 | 14 | 9 | 60.87 | Disp 1 |
| W2 | 03.09 - 03.15. | 83 | 36 | 47 | 56.63 | 23 | 13 | 63.89 | Disp 2 |
| W3 | 03.16 - 03.22. | 153 | 61 | 92 | 60.13 | 10 | 51 | 16.39 | Disp 3 |
| W4 | 03.23 - 03.29. | 137 | 56 | 81 | 59.12 | 33 | 23 | 58.93 | Disp 2 |
| W5 | 03.30 - 04.05. | 270 | 119 | 151 | 55.93 | 65 | 54 | 54.62 | Disp 4 |
| W6 | 04.06 - 04.12. | 229 | 108 | 121 | 52.84 | 53 | 55 | 49.07 | Disp 5 |
| W7 | 04.13 - 04.19. | 104 | 48 | 56 | 53.85 | 34 | 14 | 70.83 | Disp 2 |
| W8 | 04.20 - 04.26. | 206 | 87 | 119 | 57.77 | 52 | 35 | 59.77 | Disp 6 |
| W9 | 04.27 - 05.03. | 241 | 94 | 147 | 61.00 | 25 | 69 | 26.60 | Disp 3 |
| **Tot.:** | **03.02 - 05.03.** | **1475** | **632** | **843** | **57.15** | **309** | **323** | **48.89** | |

# 7 Limitations

**Input data** – only a half year log in the first period and a 9 weeks log in the second period were used. In order to gain a more holistic view (and to possibly refactor the support activities of other teams) a larger data input may be used.

**In-depth analysis, further metrics to be analysed** - there is room to define further, more complex metrics for the analysis (e.g. analysing behaviour of most active users, average response time, length of conversations etc.). However, with the scope of reducing interrupts and with the potential gain, the metrics investigated were considered sufficient.

**Changes in roles and within organisation** – a mid-size IT organisation, especially if it is transforming from a start-up to a multinational company has many changes even within a half-year period. These changes were not taken into account (e.g. there were multiple changes within the investigated team: role changes or changes among teams). Further changes occurred between the two investigated periods (e.g. the number of support chat members increased from 154 to 173 between the ends of the two investigated periods) which were not taken into the account.

# 8 Conclusion

The scope of this paper was to answer the question: "How interruptions from technical support chat could be reduced based on chat log analysis?"

In order to answer the question, 7 metrics were analysed and 7 conclusions were deducted, which helped in identifying 4 gaps serving the basis for identifying 4 action items.

The analysis showed that developers are interrupted many times during core developer hours by support chat requests. Data analysis showed that support chat interrupts at the investigated team could easily be reduced and developers could be protected by implementing action items identified in section 5, namely: (1) definition of a dispatcher service policy with a (2) (weekly rotating) dispatcher role, (3) development and maintenance of a knowledge base and (4) definition and institutionalization of a support chat mute guide.

With the analysis done and action items identified, the team started to implement the action items: they defined the dispatcher role, dispatching ideas were collected and summarized in a dispatcher policy - co-authored by the team and externals. When forming the dispatching service policy, ITIL [7] and the advantages of T-shaped people [8] were also taken into account.

After 9 weeks of operation in the new settings, support chat team members were asked to share their experiences with the new dispatcher service. Based on their feedback, it can be concluded that they experienced a clear improvement: despite the increased number of messages on the chat, their time spent on support varied between 0,5 to 8 hours per week for those weeks when they were serving as dispatchers, compared to the previous scenario when no one was clearly responsible for replying to the requests, but everyone was watching the chat and was interrupted. During the weeks investigated the team members did not need to watch the support chat (they were asked to help only in cases they were the experts of an area). The fact that ~49% of team messages were sent by dispatchers reflects that the new scenario is in place. A minor drawback is that the percentage of the messages sent by the investigated team versus all messages raised from 33 to 57.

We conclude that the way the problem was approached (dispatcher policy, dispatcher role, FAQ page and mute guide) can reduce the number of interrupts and requires a limited support time from the dispatchers. Further cases or higher amount of data may be studied before the generalisation of the approach. As a future direction, further techniques (e.g. data/text/process mining approaches, performance evaluation, formal methods, probability theory, polling systems or complexity analysis [9], [10]) and tools (such as ProM, ProM for RapidMiner or DISCO) could be involved to conduct a more detailed data analysis with the goal of understand underlying processes and analysing behaviour of support chats. Interrupts are not (and probably cannot be) fully eliminated as new problems may arise with the improvements (see section 6). As an additional future direction, the introduction of interrupt recovery techniques [11], [12] could also be investigated.

## 9 Acknowledgement

We would like to thank Nicola Grapputo for management support and András Bakonyi, Gábor Pap, Krisztina Preklet, Ferenc Tükör, Zoltán Gaál, Attila Simon and Gergely Gózca and the whole team for actively participating in the improvement project and coming up with useful ideas.

## *11 Author CVs*

**Zádor Dániel Kelemen**

Zádor Dániel Kelemen got his BSc in computer science from Gh. Asachi University of Iași, Romania, his MSc from Budapest University of Technology and Economics, Hungary and his PhD degree from Eindhoven University of Technology, Netherlands in the field of Software Process Improvement. He previously worked at SQI and ThyssenKrupp Presta. He is currently the quality manager of Common Technology Business Unit at NNG, Hungary.

**Balázs Tódor**

Balázs Tódor is the leader of the team mentioned in the article. He has an MSc in electrical engineering, a ten year long track record in game development, programming, and also some experience in leadership.

**Sándor Hodosi**

Sándor Hodosi is a project manager at NNG working in the Engine Team. He got his MSc in Computer Science at the University of Szeged. Previously he worked as a lead developer at Artifex Ltd.

**Ákos Somfai**

Ákos Somfai is a senior developer and Scrum master at the team mentioned in the paper. He received his BSc in programming and mathematics from Eötvös Lóránd Univeristy. Before joining NNG in 2010, he worked in the computer game development industry as a developer, lead developer and Scrum master at companies such as Appaloosa Interactive and Eidos Interactive Hungary.  He defined the refactor of the support chat to save more time for software development.